\documentclass[apj]{emulateapj}
\usepackage{natbib}

\usepackage{threeparttable}
\usepackage{booktabs}

\usepackage{longtable}

\usepackage{multirow}

\usepackage{amsmath}
\def \bea {\begin{eqnarray}}     
\def \ena {\end{eqnarray}}          
\def \bee {\begin{equation}}
\def \ene {\end{equation}}
\def \gas {{\rm gas}}

\def	\H		{\rm {H}}
\def	\K		{~\rm {K}}
\def	\cm		{\rm {cm}}

\def	\s		{\, \rm {s}}

\def    \apjl  		{\rm {ApJL}}
\def    \apj  		{\rm {ApJ}}
\def    \mnras  	{\rm {MNRAS}}
\def    \araa  		{\rm {ARA\& A}}
\def    \apjl  		{\rm {ApJL}}

\begin{document}

\title{Two modes of carbonaceous dust alignment}
\author{A. Lazarian}
\affil{Department of Astronomy, University of Wisconsin-Madison, USA;
\href{mailto:alazarian@facstaff.wisc.edu}{alazarian@facstaff.wisc.edu}}
\affil{Center for Computation Astrophysics, Flatiron Institute, 162 5th Ave, New York, NY 10010}

\date{Draft version \today}            

\begin{abstract}
Radiative Torques (RATs) or Mechanical Torques (METs) acting on irregular grains can induce the alignment of dust grains in respect to the alignment axis (AA), which can be either the direction of the magnetic field, or the direction of the radiation. We show that carbonaceous grains can be aligned with their axes both parallel and perpendicular to the AA and we explore the conditions where the particular mode of alignment takes place. We identify a new process of alignment of charged carbonaceous grains in a turbulent magnetized interstellar medium with respect to an electric field. This field acts on grains accelerated in a turbulent medium and gyro-rotating about a magnetic field. The electric field can also arise from the temporal variations of magnetic field strength in turbulent compressible media. The direction of the electric field is perpendicular to a magnetic field and the carbonaceous grains precess in the electric field because of their electric moments. If this precession is faster than Larmor precession in the magnetic field, the alignment of such grains is with their long axes parallel to magnetic field. We explore the parameter space for which the new mechanism aligns grains with long axes parallel to the magnetic field. We compare this mechanism with another process that provides the same type of alignment, namely, the RAT alignment of grains with insufficiently fast internal relaxation. We describe the conditions for which the particular mode of carbonaceous grain alignment is realized and discuss what information can be obtained by measuring the resulting polarization. 
 \end{abstract}

\section{Introduction}

Aligned grains both polarize starlight and emit polarized infrared radiation with the radiative torque (RAT) mechanism generally accepted to be the main driver of grain alignment (see Andersson et al. 2015 for a review). The observed polarization is correlated with magnetic fields, which makes aligned grains the major informant about magnetic fields in the diffuse interstellar medium (ISM), molecular clouds, and circumstellar regions (see Draine 2009, Crutcher 2012). The polarized dust radiation interferes with the attempts to measure polarized radiation from the Early Universe, in particular, with the search of the enigmatic B-modes of cosmological origin (BICEP2 Collaboration 2014; Kamionkowski \& Kovetz 2016; Planck Collaboration et al. 2016). To progress in studying the B-modes, the ability to interpolate dust polarization measured at one range of frequencies to another range is required. This calls for better understanding of grain alignment of dust grains of various composition.

Interstellar dust has both silicate and carbonaceous components, and it is widely believed that the two components constitute distinct populations in the diffuse ISM (see Draine \& Lee 1984). The silicate component results in polarization, while the search for polarization arising from the 3.4 $\mu$m feature arising from the aliphatic CH bond has not been observed so far in the ISM (Chiar et al. 2006),  making most of the researchers believe that carbonaceous grains are not aligned in the diffuse ISM (see Draine 2003). 

It was speculated in Lazarian et al. (2015) that the lower magnetic moment of carbonaceous grains makes it more difficult for such grains to become aligned with respect to the magnetic field and therefore to induce polarized radiation. The magnetic response of carbonaceous grains and their ability to be aligned with respect to the magnetic field were further explored in Lazarian \& Hoang (2019). There it was concluded that the composite grains containing silicate and carbonaceous fragments are expected to be aligned in the molecular clouds by the RATs, but no definitive conclusion was reached in relation to the carbonaceous grain alignment in the diffuse ISM.  

A possibility that the randomization can be the cause of the difference in the alignment of carbonaceous and silicate grains was explored in Weingartner (2006, henceforth W06).  W06 considered the randomization of such grains arising from the interaction of their time-varying electric moments with the ambient electric field. The latter arises from the turbulent acceleration of dust grains (see Yan \& Lazarian 2002).  In this paper, we subject the aforementioned physical processes. 

In this paper, we identify new effects that are induced by the electric field acting on carbonaceous grains. In particular, we consider a new process of alignment, namely, the alignment of carbonaceous dust grains in relation to the electric field. The first time, as far as we know, the alignment with respect to electric field was considered in Lazarian (2007). The source of the electric field was rather exotic, i.e. it was the field of electrostatic origin arising in the comet atmosphere. This setting was further elaborated later in Hoang \& Lazarian (2016). Here we discuss the alignment in respect to the electric field acting when grain either moves in respect to magnetic field or the value of the ambient magnetic field is changing. In particular, we show that a carbonaceous grain with velocity $V_{grain}$ gyrating around the ISM magnetic field ${\bf B}$ will experience the electric field $\sim {\bf B}\times {\bf V}_{grain}$ that can serve as the alignment axis (AA) for $10^{-5}$ cm grains moving in turbulent ISM.

 The accepted theory of grain alignment (see Andersson et al. 2015) assumes that the grains are aligned with their angular momentum {\bf J} parallel to the magnetic field {\bf B} and with {\bf J} aligned along the axis of the grain maximal moment of inertia. This constitutes the "right" alignment with the grain long axis perpendicular to {\bf B}. The far infrared polarization in this case is perpendicular to {\bf B}. This, however, is not the only outcome for the alignment. The alignment may also be also "wrong", that is, with the long grain axes parallel to {\bf B}. In this case, the far infrared polarization is parallel to {\bf B}.

We demonstrate in this paper that the new alignment process that we discuss can induce "wrong" alignment of carbonaceous grains. To simplify further searches of the carbonaceous grain alignment, we also explore the joint action of this process in parallel with another process that can also be the cause of the "wrong" alignment. In particular, we discuss the alignment of carbonaceous grains in the conditions when internal relaxation is inefficient (see Hoang \& Lazarian 2009).  In parallel to the alignment by RATs we also consider the alignment arising from Mechanical Torques (METs) that arise from the interaction of an irregular grain with a flow of atoms (Lazarian \& Hoang 2007b). 

 For both RAT and MET alignment, the alignment can happen both with respect to the magnetic field and the direction of the radiation anisotropy or the gaseous flow. We consider this type of alignmnet, the ${\bf k}$- alignment as well as the alignment with respect to {\bf B}, the ${\bf B}$- alignment (Lazarian \& Hoang 2007).

In what follows,  in \S \ref{env} we briefly introduce the environmental conditions where carbonaceous grains find themselves, while in \S \ref{sec:dyn}  we discuss  grain precession with respect to magnetic, electric fields, as well as with respect to the radiation and gaseous flow. In \S \ref{sec:align} we consider the effects of the electric field for the RAT alignment of $10^{-5}$ cm grains and alternative alignment mechanisms acting on carbonaceous nanoparticles.  The observational consequences of the predicted alignment are discussed in \S \ref{observ}. Our discussion is presented in \S \ref{sec:dis} and we summarize our findings in \S \ref{sec:sum}. 

\section{Environment of carbonaceous grains}
\label{env}

\subsection{Grain charging}
\label{charging}

 In general, in typical astrophysical environments, in particular, in interstellar environments, carbonaceous grains are charged. The grain charging can arise from ions/protons colliding with the grain surface. If the temperatures of electrons and ions are the same, the electrons have higher velocity than the ions. As a result, the electrons collide more frequently with the grain surface compared to ions and the grain gets an excessive negative charge. The corresponding electric potential slows down the electrons to make the rate of their collisions the same as ions. Naturally, the higher the temperature of the gas, the higher the collisional negative charging of grains.  

Electron photoemission is a competing process that tends to make the grain charge positive. Indeed, the energetic photons tend to remove electrons from the grains. 

In most astrophysical environments, both processes are acting together, competing with each other. The details of both mechanisms are not trivial and require further investigation. For instance, for the process of collisional charging, the capturing electrons may depend on the process of energy transfer from impinging electrons to the grain material. The process of photoemission depends on the photoelectric yield of the grain, which varies significantly. The charging for nanoparticles (see Draine \& Lazarian 1998) is different from the charging for classical grains.  The work function of different carbonaceous materials depends on the variations of grain composition as well as the the grain environment. More details of these complex processes can be found in a number of dedicated studies (see Weingartner \& Draine 2001, see also Draine 2011). This paper outlines the general peculiarities of the alignment of carbonaceous grains, and the discussion of the  particular regimes of charging of carbonaceous grains is beyond its scope.  Instead, for the estimates in this paper we use the characteristic values from the available literature.

Within our study, we use the estimate of grain charge to evaluate the rate of grain Larmor rotation about a magnetic field, and we use the grain dipole moment estimates from Weingartner (2009) and Jordan \& Weingartner (2009). In future publications, we intend to address the modification of alignment processes for carbonaceous grains of different sizes and composition by the variations of grain charging, as well as the variations in the grain electric dipole moment (see Appendix A).

\subsection{Grain acceleration}
\label{acc}

Different processes can accelerate charged grains, the radiation pressure (see Purcell 1969) and shocks (see Zhu et al. 2019) being frequently discussed. We, however, focus on turbulence as the mechanism for increasing $V_{grain}$. 

Observations of the ISM testify that this medium is turbulent (see Armstrong et al. 1995, Chepurnov \& Lazarian 2010, McKee \& Ostriker 2007). Turbulence accelerates interstellar grains and is the main driver for the $10^{-5}$~cm grains in typical interstellar conditions
(Lazarian \& Yan 2002, Yan \& Lazarian 2003, Yan et al. 2004, Hoang et al. 2012). The velocities of charged grains vary depending on the level of turbulence and grain size. For our estimates, we will assume that $10^{-5}$ cm grains can get velocities of the order of 1 km/s. 

 We note that for the present study the motion of grains perpendicular to the magnetic field is important. For instance, the radiation pressure can efficiently accelerate grains along the magnetic field, while the grain gaining momentum perpendicular to the magnetic field direction is suppressed by the effect of the Larmor force. The scattering of charged grains by turbulent magnetic fluctuations, however, can transfer the grain momentum parallel to the magnetic field into the perpendicular one. Turbulence can induce additional effects; for instance, turbulent magnetic reconnection (Lazarian \& Vishniac 1999) can induce the acceleration of charged grains, similar to the acceleration of energetic particles (see Kowal et al. 2011). 

Apart from turbulence, shocks can be an important process that locally can move grains perpendicular to the magnetic field. The corresponding grain-gas velocities in shocks may be sufficiently large to make the METs (see Appendix C) stronger than the RATs. In addition, the ambipolar grain drift in cores of molecular clouds can be supersonic according to Roberge et al. (1995).

\subsection{Radiative and Mechanical torques}

RATs  (Dolginov \& Mytrophanov 1978, Draine \& Weingartner 1996, 1997, Lazarian \& Hoang 2007a, henceforth LH07) produce grain alignment, and several regimes of such alignment exist. As discussed in more detail in Appendix B, the alignment can happen with low and high angular momentum at the state of alignment. The Analytical Model (AMO) introduced in LH07 is equally applicable to carbonaceous and silicate grains. Therefore, one expects that the alignment of both types of grains to be present in the ISM, provided that both types of grains have similar sizes. In the paper below we discuss the peculiarities that, nevertheless, are present in the process of the carbonaceous grain alignment. 

METs were introduced in Lazarian \& Hoang (2007b) in analogy with RATs, appealing to the same model of a helical grains. The only difference is that  atoms rather than photons are  interacting with the irregular grain to induce METs. More details about METs are summarized in Appendix C. While the RAT theory is well tested with extensive polarimetric observations (see Andersson et al. 2015), the cases of MET alignment have not been observationally identified yet. However, the support of the importance of METs comes from numerical calculations of torques obtained by exposing irregular grains with a flow of bombarding atoms in  Hoang et al. (2018)
as well as calculations in Das \& Weingartner (2016). In typical interstellar conditions, the METs can be subdominant to RATs, but they have their own niche. For instance, METs can align grains in the regions of reduced radiative flux as well as grains with size $a_{eff}$ much less than the typical radiation wavelength $\lambda$ present in the grain environment.

\section{Precession of carbonaceous grains} 
\label{sec:dyn}

\subsection{Precession in electric field in turbulent media}
\label{sec:turb}

Electric field effects are usually disregarded in discussing grain dynamics. The exception is the small grains in the ISM that have a Debye scale larger than the mean distance between the grains. Such grains, as shown in Hoang \& Lazarian (2012), can be accelerated due to the variations of their charge. We do not consider electrostatic electric fields of this type or similar electrostatic fields in the atmospheres of astrophysical objects, e.g. comets (Lazarian 2007, Hoang \& Lazarian 2016). Instead, we consider the electric field arising from the change of magnetic field in the grain reference frame or the grain motion with respect to the magnetic field. This field and the  in it were  for the first time, as far as we know, considered in  W06 in the framework of the mechanism of randomization of grains that we discuss in \S \ref{sec:rand}.  In this paper, we focus on the other effects induced by the motion of the charged grain.  

Consider the simplest setting of a charged grain performing gyrorotation about the direction of a magnetic field ${\bf B}$ as it illustrated in Figure 1. The electric field ${\bf E}_{induced}$ acting on this grain is $\sim {\bf V}_{grain} \times {\bf B}$. In the electric field ${\bf E}_{induced}$ the grain with electric moment $\mu_q$ performs a precession $\sim \mu E_{induced}$. The corresponding precession for carbonaceous grains, as we discuss below, can be much faster than the precession of a grain in magnetic field. This is the consequence of the of carbonaceous grains having a relatively small magnetic moment. As a result, the electric field can define the AA for carbonaceous grains.  

\begin{figure}[htbp]
\centering   
   \includegraphics[width=9cm]{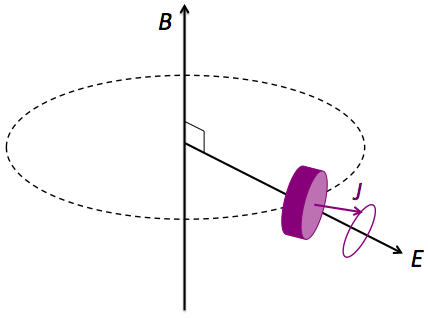}
\caption{As a charged grain gyrorotates about a magnetic field it experiences electric field ${\bf E}$. This field induces the precession of grain angular momentum ${\bf J}$.  If $\Omega_E>max[\Omega_{B}, \Omega_k]$ the  alignment happens with respect to ${\bf E}$, which is perpendicular to ${\bf B}$. In the presence of fast internal relaxation, grain long axes are parallel to ${\bf B}$.}
\label{E_alignment}
\end{figure}

A charged grain moving with respect to magnetic field, for example, gyrating about a magnetic field,  experiences the electric field with amplitude
\begin{equation}
E_{induced}=\frac{V_{grain,\bot}}{c} B,
\label{induced}
\end{equation}
where $V_{grain,\bot}$ is the perpendicular to the magnetic field component of grain velocity. If
the grain has an electric dipole moment component parallel to its angular momentrum ${\bf p}_{el,J}$, then the electric torque acting on the grain is
\begin{equation}
{\bf \Gamma}_{el}={\bf p}_{el,J}\times {\bf E}_{induced},
\end{equation}
which induces precession about the electric field direction.

The rate of the precession in an electric field should be compared with the rate of Larmor precession in the magnetic field:
\begin{equation}
\Omega_{\rm L}=\frac{\mu B}{I_{\|}\omega},
\label{eq:tl}
\end{equation}
where $I_{\|}$ is the grain maximal moment of inertia. 

For a charged grain,
the 
 corresponding magnetic moment is
\begin{equation}
\mu_{q}=\alpha \frac{qJ r^2}{3cI_{\|}},
\end{equation}
where $J$ is grain angular momentum, $r$ is the effective radius of the grain, and $\alpha$ is the parameter responsible for the charge distribution. Using the values in adopted in W06 for the cold neutral medium (CNM), one gets
\begin{equation}
\label{charge}
\mu_{\rm charge} \approx 1.1 \times 10^{-24} \, 
U_{0.3} a_{-5}^2 \left( \frac{\omega}{10^5 \s^{-1}} \right)
\, {\rm stat~coulomb~cm},
\end{equation}
where $U_{0.3}=(U/0.3V)$ is the grain potential normalized by the electric potential expected for a $10^{-5}$ cm grain in the CNM, $a_{-5}$ is the grain size normalized by $10^{-5}$cm, and $\omega$ is grain velocity. This is the lower limit of the grain magnetic moment.

Nuclear spins of hydrogen atoms in carbonaceous grains can result in the rotating grain gaining its magnetic moment through the Barnett effect (see Dolginov \& Mytrophanov 1976, Purcell 1979). Adopting the assumption in W06 that hydrogen constitutes 10\% of atoms in a carbonaceous grain, one gets a higher magnetic moment
\begin{equation}
\label{Bar_carb}
\mu_{\rm Bar} 
\approx 1.5 \times 10^{-23} \, T_{15}^{-1} a_{-5}^3 \left( \frac{\omega}
{10^5 \s^{-1}} \right) \, {\rm stat~coulomb~cm},
\end{equation}
where grain temperature $T_{15}=T/(15K)$ is normalized to 15K.

Lazarian \& Draine (2000) employed higher  magnetic response of carbonaceous grains appealing to the existence of unpaid bonds and free radicals.  W06 argued, however, that the value of for the electron spin density should be $\sim 10^{19} {\rm g}^{-3}$ which agrees with the results for hydrogenated amorphous carbon in Esquinazi \& Hohne (2005). In the latter case, the magnetic moment will still be an order of magnitude larger than the estimate given by Eq. (\ref{Bar_carb}).   

In the external magnetic field $B$, the precession rate of a grain 
\begin{equation}
\label{prec}
\Omega_B  \approx 0.02\bar{\mu} a_{-5}^{-5} \left( \frac{B}
{5 \, \mu {\rm G}} \right) \left( \frac{\rho}{3 {\rm g/cm}^{-3}} \right)^{-1}
 \left( \frac{\omega}
{10^5 \s^{-1}} \right)^{-1} \, {\rm yr}^{-1},
\end{equation}
where $\bar{\mu}$ is normalized by the value of the Barnett moment given by Eq. (\ref{Bar_carb}).

The rate of the precession  in an  electric field should be compared with the previously discussed Larmor precession rate:
\begin{equation}
\Omega_{E, grain}=\aleph_{grain} \Omega_B.
\label{Omegael}
\end{equation}
Here, 
\begin{equation}
\aleph_{grain}=\frac{p_{el, J}V_{grain, \bot}}{\mu c} ,
\label{aleph0}
\end{equation}
and the component of the grain electric moment projected on the direction of grain angular momentum ${\bf J}$ is
\begin{equation}
\label{pel}
p_{el}\approx 1.0 \times 10^{-15} U_{0.3} a_{-5}^2 \left( \frac{\kappa_{el}}
{10^{-2}} \right) {\rm stat~coulomb~cm}, 
\end{equation}
where $\kappa_{el}$ is a parameter describing the charge distribution. 
Due to the rapid grain rotation, the effect of other components of the grain electric moment is averaged out.

The estimate for $\aleph_{grain}$ for hydrogenated carbonaceous grains is
\begin{equation}
\label{aleph_carb}
\aleph_{grain} \approx 405 
a_{-5}^{-1} T_{15} U_{0.3} V_{grain,1}\left( \frac{\omega}{\omega_d} 
\right)^{-1}  
\left( \frac{\kappa_{el}}{10^{-2}} \right).
\end{equation}
where the grain velocity $V_{grain,1}=V_{grain}/({\rm 1~ km/s})$ is normalized by 1 km/s (see Yan et al. 2004), and $\omega_d$ is the angular velocity corresponding to the dust temperature $T_{15}$, i.e. $\omega_{d}=\sqrt{2k T/I_{\|}}$, which is  $\sim 2.1 \times 10^4$ for $T=15K$. 

We note that $\Omega_{E, grain}$ is significantly larger for $10^{-5}$~cm grains than the rate of grain gyrorotation\footnote{The details of alignment vary depending on whether $\omega_{gyro}$ is larger or smaller than $\Omega_B$ as well as inverse of grain damping time. We do not discuss those in this paper.}:
\begin{equation}
    \omega_{gyro}\approx 
    0.004~a_{-5}^{-2} U_{0.3} 
    \Big(\frac{\rho}{3 \text{g cm}^{-3}} \Big)^{-1} 
    \Big( \frac{B} 
{5 \, \mu {\rm G}} \Big)~{\rm yr}^{-1}.
\label{gyro}
\end{equation}
This justifies the schematic of grain dynamics illustrated by Figure~1. 

The turbulent variations of magnetic field strength can also be the reason for the electric field being present in the grain system of reference. It is worth noting that for the changes of magnetic field, the media does not require the media to be compressible. Even in incompressible medium one can distinguish two types of modes, the Alfven mode and pseudo-Alfven one, the latter being the limiting case for the slow modes (see discussion in Cho \& Lazarian 2003). For instance, the pseudo-Alfven mode provides magnetic field compressions that induce the electric field parallel to magnetic field with the magnitude
\begin{equation}
{\bf E}_{turb}\approx -{\bf v}_{l, turb}\times {\bf B},
\label{Eturb}
\end{equation}
where $v_{l,turb}$ the turbulent velocity at scale $l$. For the interstellar turbulent cascade, most of the energy injection takes place at large scales, and
the amplitude of velocities $v_{turb}$ is the largest at the large scales. This is true also for the slow and fast modes that compress magnetic fields in 
compressible interstellar medium (see Cho \& Lazarian 2003).  Therefore, velocities that enter Eq. (\ref{Eturb}) can be larger than the grain velocities $V_{grain}$
making $E_{turb}>E_{induced}$. As a result, the grains precess in this electric field with the rate
\begin{equation}
\label{Omega_turb}
\Omega_{turb} \approx \aleph_{turb} \Omega_B
\end{equation}
where
\begin{equation}
\aleph_{turb}==\frac{p_{el, J}v_{turb}}{\mu c} ,
\label{aleph2}
\end{equation}
and the precession rate $\Omega_{turb}$ can potentially be up to order of magnitude larger than the precession rate $\Omega_{el, grain}$ given by Eq. (\ref{Omegael}). For the rest of the paper, we will use the precession rate in the electric field given by
\begin{equation}
    \Omega_{E}=\aleph \Omega_B,~~~~{\rm where}~~~~\aleph=max[\aleph_{grain}, \aleph_{turb}].
    \label{Omegaelf}
\end{equation}

In addition, we would like to mention that $\sim 10^{-5}$~cm dust grains can reach the drift velocities $\sim 0.3$ km/s due ambipolar drift (Roberge et al. 2004). This is an important extra process to consider if carbonaceous grains can be observed in molecular clouds, especially in the environments with relatively low level of turbulence. 

We stress that in all of the cases discussed above, the electric field ${\bf E}$ 
that grains experience is perpendicular to the magnetic field. This, as we discuss in \S \ref{sec:align}, can result in direction of grain alignment being 90 degrees different from the generally accepted alignment with long grain axes perpendicular to ${\bf B}$.

 The direction of the electric field {\bf E} in the settings above is perpendicular to the magnetic field {\bf B}. Therefore, it is the magnetic field that acts as the AA in the absence of other factors inducing faster precession. For carbonaceous grains the alignment can happen with grain axes parallel or perpendicular to ${\bf B}$ depending on the relation between $\Omega_B$ and $\Omega_E$.

\subsection{Comparison with the precession induced by RATs and METs}
\label{sec:precess}

We explain in \S \ref{sec:align} that the grain alignment can take place with respect to the direction of radiation provided that the rate of precession about this direction is faster than both $\Omega_{el}$ and $\Omega_B$. The corresponding rate is given by (see LH07) 
\begin{eqnarray}
\Omega_{rad, p}&\approx& 0.057\ \hat{\rho}^{-1/2}\hat{s}^{1/3}a_{-5}^{-1/2}\hat{T_d}^{-1/2}\\
&\times&\left(\frac{u_{\rm rad}}{u_{\rm ISRF}}\right) \left(\frac{\lambda}{1.2\ \mu{\rm m}}\right) \left(\frac{\gamma \overline{|\mathbf{Q_{\Gamma}}|}}{0.01}\right)
\left(\frac{\omega_d}{\omega}\right) \ {\rm yr}^{-1},\nonumber  
\label{rad_p}
\end{eqnarray}
where $\omega_d$ is the angular velocity corresponding to the temperature of dust $T_d$, as $\omega_{d}=\sqrt{2k T_{\rm d}/I_{\|}}$  and the amplitude of RATs  $\mathbf{Q_{\Gamma}}$ is estimated from numerical calculations (LH07) as 
\begin{eqnarray}
|\mathbf{Q_{\Gamma}}|&\approx& 2.3\left(\frac{\lambda}{a_{\rm eff}}\right)^{-3}\ \ {\rm for\ } \lambda>1.8 a_{\rm eff}\\
&\approx& 0.4\ \ \ \ \ \ \ \ \ \ \ \ \ {\rm for\ } \lambda\leq1.8 a_{\rm eff} \label{eq:QRATs}.
\end{eqnarray}
where  $u_{\rm \lambda}$ is the energy spectrum of the radiation field, $\lambda$ is the radiation wavelength, and $\gamma_\lambda$ is the anisotropy parameter.  The scaling of the RAT strengths for $\lambda \ll a_{eff}$ is different and requires further studies.\footnote{ We expect to see a similarity between the RATs in the $\lambda \ll a_{eff}$ regime and the METs. This is due to the fact that, similar to an atom, a photon in this regime interacts only with a small portion of the grain and therefore the grain effective helicity is changing as the grain turns its facets to the anisotropic radiation flux.}

In addition, the realistic irregular grain moving with respect to the gas experiences METs. These torques induce the grain precession (LH07, Lazarian \& Hoang 2019):
\begin{equation}
    \Omega_{mech,p}\approx  0.009\left(\frac{\omega_d}{\omega}\right)V_{grain,1}\left(\frac{s}{0.5}\right)^{2}\sin 2\Theta ~{\rm yr}^{-1},
    \label{t_mex}
\end{equation}
where $s=a/b<1$ is the axial ratio of oblate grains, $V_{grain,1}$ is the velocity of the grain normalized by 1 km/s, $\Theta$ is an angle between the grain symmetry axis and the vector ${\bf V}_{grain}$, and $v_{th}$ is the thermal velocity of hydrogen. Turbulence tends to accelerate grains perpendicular
to the magnetic field (see Yan \& Lazarian 2003), while the radiation pressure,  as we discussed in \S \ref{acc}, 
accelerates them mostly along magnetic field. In both cases, the 
factor $\sin 2\Theta$ in Eq.(\ref{t_mex}) decreases the rate of mechanical precession.

The calculations of the precession in Eqs. (\ref{t_mex}) and (\ref{rad_p}) are performed for a slow grain rotation that corresponds to the state of the alignment at the low-J attractor point (see LH07) where the grains are most susceptible to the precession induced by external torques. If grains rotate faster, as we discuss in \S \ref{supra}, the relative role of Larmor precession is increasing.    

For the parameters adopted, comparing Eq. (\ref{rad_p}) with Eqs. (\ref{prec}), and (\ref{Omegael}), (\ref{aleph_carb})  one can observe that  $\Omega_B<\Omega_{rad, p}<\Omega_{E}$. The fastest precession determines the axis about which the alignment takes place (see LH07). The dominance of $\Omega_{E}$  which means that the alignment happens in terms of the electric field. However, as the grains perform the gyrorotation given by Eq. (\ref{gyro}) the total alignment is happening with respect to the magnetic field and the magnetic field acts as the axis of alignment.

The above relation between the precession rates is not universal. For instance, in the presence of the radiation much higher than the interstellar one, the RAT-induced precession can be much faster than the value in Eq. (\ref{rad_p}). Then, in the vicinity of stars, the relation between the angular velocities can be 
$\Omega_B<\Omega_{E}<\Omega_{rad, p}$, that is, the alignment is happening with respect to the radiation direction. For grains of sizes significantly smaller than the typical wavelength of the impinging radiation, the RATs can get subdominant. This is the case, for instance, for the carbonaceous nanoparticles that we discuss in \S \ref{nano}. The thermal velocities of such grains are large enough to make the $\Omega_B>\Omega_E$. As we discuss in \S \ref{supra} grains can reach velocities one or two orders of magnitude larger than that assumed in Eq. (\ref{t_mex}), which can also may make $\Omega_B>\Omega_E$.

\subsection{Thermal and Superathermal rotation of dust grains}
\label{supra}

Thermal rotation of dust grains that arises from gas bombardment was considered from the very beginning of the grain alignment research (see Davis \& Greenstein 1951). The suprathermal rotation arising from uncompensated torques was introduced in the field by Purcell (1979). The three causes of the fast rotation identified there are (1) ejection of nascent H$_2$ molecules formed on grain catalytic sites, (2) photoemission of electrons from the grain surface, and (3) variations over the grain surface of the accommodation coefficient for the bombarding atoms. It was argued in Purcell (1979) that these processes can increase the grain rotation velocity $\omega$ by two orders of magnitude. 

Independent of the processes introduced by Purcell (1979), the RATs can induce high velocities of grain rotation (Dolginov \& Mytrophanov 1976, Draine \& Weingartner 1996, 1997). This happens when grains become aligned at the attractor corresponding to the high angular momentum, that is, at the so-called  high-J attractor point the parameter space for this was established is defined in LH07. The corresponding rotational velocities of $10^{-5}$~cm grains at high-J attractors in the typical interstellar radiation field  may be two orders of magnitude higher than the velocities at the low-J attractor points. The abundance of such grains is limited by the parameter space for which the grains exhibit high-J attractor point (see Figure 24 in LH07) and the time that is available for the phase-space diffusion from the low-J attractor point to the high-J attractor point (Hoang \& Lazarian 2008).\footnote{It is shown in Hoang \& Lazarian (2008) that, in the case when the grain with the high-J attractor point gets illuminated, initially, most grains follow in phase space to the low-J attractor points. In the presence of randomization arising, for instance, from gaseous bombardement, grains diffuse on several gaseous damping times to the high-J attractor.} 

The numerical analysis in Herranen et al. (2019) shows that for a significant percentage of grains in a given interstellar volume, only low-J attractor points will be present. It is interesting that for grains rotating at the low-J attractor, the Purcell torques and RATs act in different direction (Hoang et al. 2008). Due to this effect grains do not reach the rates that they would obtain with Purcell's torques acting alone. This effect decreases the degree of suprathermallity of carbonaceous grains. Thus we expect a significant part of carbonaceous interstellar grains to rotate with velocities not much larger than the thermal ones.

\section{Alignment of carbonaceous grains}
\label{sec:align}

The generally accepted picture of alignment (see Lazarian 2007) includes the alignment of grain axes with respect to its angular momentum ${\bf J}$ (internal alignment) and the alignment of ${\bf J}$ with respect to either  the external magnetic field ${\bf B}$ or the direction of radiation given by vector ${\bf k}$ (see LH07).  Similarly, in the case of the METs, depending on the relative rates of precession $\Omega_B$ and $\Omega_{rad,P}$ either the magnetic field ${\bf B}$ or the direction of the flow ${\bf k}$ provides the axis of alignment. Whether the internal alignment is efficient depends on both the rate of internal alignment and the ratio of the internal grain temperature to the effective temperature associated with grain rotation (Lazarian 1994). For instance, the internal alignment of grains rotating at the low-J attractor point is low as the grain rotational temperature is of the order of the grain temperature.
Naturally, the low internal alignment decreases the degree of polarization arising from aligned grains. A more subtle but important effect discovered in Hoang \& Lazarian (2009) is that the RAT alignment of grains on the time scales at which internal relaxation is inefficient can  turn 90 degrees to the direction corresponding to the RAT alignment in the presence of internal relaxation. As it described in Appendix B, for the stabilization of the alignment in this position one requires the action of Purcell (1979) pinwheel torques. For our considerations below we assume that these torques are present. In other words, the RAT or MET alignment of grains in the case of inefficient internal relaxation, similar to the case of the alignment in relation to electric field, can induce a 90 degrees change from the "classical" expectations of grain alignment (see Andersson et al. 2015).   

\subsection{Internal relaxation in carbonaceous grains}
\label{intern}

The alignment of grains critically depends on the efficiency of internal relaxation within the grain material. The relaxation takes place as a grain
wobbles about the direction of its angular momentum ${\bf J}$ with inelastic relaxation being the simplest relaxation process (Purcell 1979).  For instance, for an oblate graphite grain, Lazarian \& Efroimsky (1999) obtained
\begin{equation}
\tau_{inelast}\approx 2.3\times 10^{-1} a_{-5}^{5.5} X_{10}^{3/2} \left( \frac{\rho}{3 \text{g cm}^{-3}} \right)^{1/2} {\rm yr}, 
\end{equation}
where $X_{10}$ is the grain axis ratio normalized by 10. 

Lazarian \& Draine (1999) identified a significantly faster mechanism of internal relaxation related to nuclear spins. The time scale for the nuclear relaxation for a "brick" with dimensions $a\times a\sqrt{3}\times a\sqrt{3}$ is
\begin{equation}
\tau_{\rm NR}\approx G_{\rm NR} \rho_{-3}^2 a_{-5}^{7},
\label{NR}
\end{equation}
where
\begin{equation}
G_{NR}=0.8 \times 10^{-5} \left(\frac{n_e}{n_n}\right) \left(\frac{\omega_d}{\omega}\right)^2 \hat{\mu_N} \left[1+\left(\omega \tau_n \right)^2 \right] {\rm yr}, 
\label{Gn}
\end{equation}
where $\hat{\rho}\equiv \rho/(2 {\rm g~cm}^{-3})$, $\omega_d$ is the angular velocity of a grain rotating thermally at $T_d=10$K, that is, $\approx 3\times 10^4$ s$^{-1}$; $\hat{\mu_N}$ denotes the nuclear magnetic moments $\mu_n=g_{n}\mu_{N}$ normalized by the magnetic moment of the proton $\mu_N\equiv 5.05 \times e\hbar/2m_pc=10^{-24}$ erg G$^{-1}$; and $\tau_n$ is the time of spin-spin relaxation within the system of nuclear spins:
\begin{equation}
\tau_n\approx 0.18\times 10^{-3} \left(\frac{10^{22} cm^{-3}}{n_n}\right)~s
\end{equation}
which for slow grain rotation provides $\tau_{NR}\ll \tau_{inelast}$, but becomes comparable and longer for grains rotating with suprathermal angular velocities $>10^2\omega_d$. 

The internal relaxation rate becomes
\begin{equation}
    \tau_{int}^{-1}=\tau_{inelast}^{-1} + \tau_{NR}^{-1},
    \label{int_tot}
\end{equation}
when two mechanisms work simultaneously. We have not included in our discussion of internal relaxation the Barnett relaxation (Purcell 1979), as this type of relaxation is significantly reduced for carbonaceous grains due to the low density of unpaired electron spins. For the sake of simplicity, we also disregarded additional internal relaxation mechanisms introduced in Lazarian \& Hoang (2019). Those do not appreciably change the overall relaxation rate given by Eq. (\ref{int_tot}).

\subsection{RAT and MET alignment}

The RAT alignment (Dolginov \& Mytrophanov 1976, Draine \& Weingartner 1996, 1997, LH07) is a theoretically understood process. The Analytical Model (AMO) introduced in LH07 has been shown to reproduce well the properties of RATs acting on various irregular grains (see Herranen et al. 2019, Herranen \& Lazarian 2020). The alignment by regular METs acting on irregular grains was introduced in Lazarian \& Hoang (2007b) in analogy with the AMO. The importance of the METs and the efficient grain alignment  by these torques were confirmed in the subsequent studies (Das \& Weingartner 2016,
Hoang et al. 2018). While for the RATs the AMO provides a quantitative description of the torque dependences, the METs happen to be more complex. Therefore, the AMO provides only the qualitative description of the helicity effects for the METs. In what follows, we, nevertheless, use the  numerically confirmed general analogy between the action of the RATs and METs (see Hoang et al. 2018, also Appendix B). to present their action on carbonaceous grains within our unified approach.  

For both the RAT and MET alignment, the rate of internal relaxation plays a very important role. The theory of grain alignment usually assumes that the time of internal alignment is shorter than the RAT or MET alignment times. This is the most common case for aligned grains in a diffuse ISM, but it is not the only possible setting. As we mentioned earlier, if the alignment happens on the time less than the internal relaxation time, then the properties of alignment may be very different, as shown in Hoang \& Lazarian (2009).

\subsubsection{Fast Alignment without internal relaxation}
\label{FA}

Consider first the case of the Fast Alignment (FA), that is, the alignment that happens in the radiative flux on the time scales much less than the gaseous damping time. This type of alignment can take place in the presence of time-variable radiation sources, such as, novae, supernovae, and variable stars. For an intensity of radiation 100 times the averaged interstellar one, the FA takes place for grains larger than $0.6\times 10^{-5}$~cm (Hoang \& Lazarian 2009). 

The FA introduced in LH07 takes place on the timescale $\Omega_k^{-1}$, where $\Omega_k$ is $max[\Omega_{rad,p}, \Omega_{mech, p}]$, where the precessions in the radiative and mechanical fluxes are given by
Eqs. (\ref{rad_p}) and (\ref{t_mex}) respectively.  

Hoang \& Lazarian (2009) considered the alignment with respect to the magnetic field. If the timescale of FA is shorter than the time scale of internal relaxation for carbonaceous grains, the results are summarized in Table 1 in Hoang \& Lazarian (2009). They testify that such grains at low-J attractor points become aligned with long axes parallel to the magnetic field, assuming that $\Omega_B$ is the fastest precession rate. If $\Omega_k>\Omega_B$ the expected alignment is with long grain axes parallel to the radiation direction or the direction of the mechanical flow. 

If in the process of FA the grains reach the high-J attractor points, they become aligned with the long axes perpendicular to either ${\bf B}$ or ${\bf k}$, whichever is the direction of the fastest precession.
Figure 24 in LH07 defines the parameter space for the existence of high-$J$ attractor points in terms of the grain parameter $q^{max}$ and the angle $\psi$ between the grain axis of alignment and the direction toward the radiation source. It testifies that for a significant part of of the parameter space the grains have only low-J attractor points. As we mentioned before (see Hoang \& Lazarain 2008), even in the presence of the high-J attractor point, most of the grains stream in the $[q^{max}, \psi]$ phase space to the low-J attractor point.
Thus, the FA of grains in the absence of internal relaxation results in grains aligned  with their long axes parallel to the direction of its maximal precession rate. The grains aligned in this way can produce, for instance, far infrared linear polarization parallel to ${\bf B}$.

In this paper, we have discussed grain precession with respect to ${\bf E}$ and argued in \S \ref{sec:dyn} that for $10^{-5}$~cm carbonaceous grains $\Omega_E$ can be expected to be larger than $\Omega_B$. This difference, however, does not change the nature of the RAT or MET alignment. If $\Omega_{E}>\Omega_k$ the FA alignment discussed above will happen with long axes parallel to the electric field ${\bf E}$. Evidently this type of alignment will result in grains being aligned with long axes {\it perpendicular} to magnetic field ${\bf B}$. In other words, the two effects, namely, the one of fast ${\bf E}$ precession and the FA in the presence of the inefficient internal relaxation, can {\it compensate} for each other. As a result, the polarization arising from the fast alignment of carbonaceous grains may become similar to the polarization arising from the alignment of silicate grains.

\subsubsection{Intermediate and stationary alignment without internal relaxation}
\label{IA}

The minimal randomization time for a grain is the gaseous damping time of the grains, which is 
\begin{eqnarray}
    \tau_{\rm rand}&\approx &\frac{3}{4\sqrt{\pi}}\frac{I_{\|}}{n_{\H}m_{\H}
v_{\rm th}a^{4}} \nonumber\\
&=&7.3\times 10^{4} a_{-5}\left( \frac{\rho}{3 \text{g cm}^{-3}} \right) \left(\frac{100\K}{T_{\gas}}\right)^{1/2}\left(\frac{30
\cm^{-3}}{n_{\H}}\right) {\rm yr},~~~~\label{eq:taugas}
\end{eqnarray}
where $n_H$ is the density of ambient hydrogen gas with temperature $T_{gas}$. In the diffuse ISM, in the presence of the infrared emission and other damping (see Draine \& Lazarian 1998) the resulting damping time can be several times smaller that that given by Eq. (\ref{eq:taugas}). The latter estimate also drops as $1/n_H$ as the density of gas increases. At the same time, the internal relaxation time increases as $a^7$ for nuclear relaxation and $a^{5.5}$ for inelastic relaxation. Therefore, if large carbonaceous grains can exist in the dense environments of molecular cloud cores and accretion disks (see Perez et al. 2015), such grains can have the internal relaxation time $\tau_{int}$ less than $\tau_{ran}$ (see more in \S \ref{observ}). In this setting, the alignment in the absence of internal relaxation can happen in the stationary radiation field or mechanical flow.  As we discussed earlier, if grain rotation is stabilized at the low-$J$ attractor points by Purcell (1979) pinwheel torques, the resulting alignment is "wrong" in its direction, that is, long grain axes are aligned parallel to the AA.

The grain alignment with inefficient internal relaxation is different for the parameter space where grains have and do not have high-J attractor points (see LH07). For grains with only low-$J$ attractors, the long grain axes being parallel to the axis of maximal precession rate (henceforth, axis of maximal precession) is only possible. If, however, a high-$J$ attractor exists for a given grain within a $[q^{max},\psi]$ parameter range, the initial alignment with long grain axes parallel to the maximal precession axis will gradually evolve. It will change to the alignment with grain axes perpendicular to this axis on the time scale of the order of several $\tau_{rand}$. The latter follows from the calculation in Hoang \& Lazarian (2008) of the evolution of grain phase trajectories in the presence of gaseous bombardment. 

On the basis of the discussion above, we can introduce for grains with high-J attractors and with $\tau_{int}>\tau_{rand}$ two new types of alignment. One is Intermediate Alignment (IA) that happens with long grain axes parallel to the maximal precession axis, for example, ${\bf B}$ or ${\bf k}$. This anomalous alignment is important on the time scales up to $\sim \tau_{rand}$. The Stationary Alignment (SA) that takes place over time $\gg \tau_{rand}$ is different and is characterized by the long grain axes alignment perpendicular to the maximal precession axis. This change of the direction of alignment taking place on $\sim \tau_{rand}$ time in the vicinity of variable radiation sources can be observable.

If grains correspond to the parameter space $[q^{max},\psi]$ with only low-J attractor point, there is no change of the  alignment with time. In other words, both IA and SA correspond to the anomalous alignment with long grain axes parallel to the direction of maximal precession. 

In terms of polarization, it is accepted that, for silicate grains with efficient internal relaxation, the RAT or MET alignment provides  far infrared polarization that is perpendicular to either to ${\bf B}$ or ${\bf k}$, depending on whether $\Omega_B$ of $\Omega_k$ is larger (LH07). For carbonaceous grains, more possibilities are present. 
Depending on the relation between $\Omega_k$, $\Omega_B$ and $\Omega_{el}$, ${\bf k}$ or ${\bf E}$ or ${\bf B}$ correspond to the direction of maximal precession. The anomalous alignment is with the long grain axes parallel to ${\bf k}$ if $\Omega_k>max[\Omega_B, \Omega_E]$. $\Omega_B>\Omega_{E}$ the polarization from carbonaceous grains is parallel to ${\bf B}$ and if $\Omega_{E}>\Omega_B$ it is perpendicular to ${\bf B}$. The polarization direction flips 90 degrees on the timescale $\sim\tau_{rand}$ if grains have high-J attractors.

\subsubsection{Carbonaceous grain alignment: two modes}

Traditionally, the grain alignment theory was developed to explain the alignment of silicate grains. These grains are characterized by an abundance of uncompensated electron spins, which results in both in a higher grain magnetic moment $\mu$ and also efficient relaxation arising from the Barnett effect (Purcell 1979). Pure carbonaceous grains with size $\sim 10^{-5}$~cm have significantly smaller $\mu$ and have less efficient internal relaxation. 

As a result of these differences, two effects that are mostly disregarded within the accepted grain alignment theory become important. The most important is the new effect of the alignment in relation to the electric field. The other effect is the alignment of carbonaceous grains in the absence of internal relaxation. While the latter process was addressed earlier in Hoang \& Lazarian (2009), the case of the carbonaceous grain alignment demonstrates new facets of the process.

Our results are summarized in Table 1 which presents the conditions for the alignment of carbonaceous dust parallel and perpendicular to the magnetic field ${\bf B}$ and the radiation direction or gaseous flow
${\bf k}$. We consider both stationary and time-dependent alignment. Thus, $\tau$ is the time for the alignment in the case of time-dependent radiation or mechanical flux,   

Table 1 demonstrates the bimodal nature of alignment with respect to ${\bf B}$. The results show that the interpretation of polarization measurements based on the generally accepted understanding of grain alignment theory can be highly misleading as far as the alignment of carbonaceous grains is concerned. The fact that the alignment can be both with respect to both ${\bf B}$ and ${\bf k}$ is well accepted in the community. This was discussed in detail back in LH07 for the case of RAT alignment, and the arguments there are directly applicable to the MT alignment (Lazarian \& Hoang 2007b). Thus in Table 1 both processes are discussed together and $\Omega_k$ is the maximal precession rate induced by the RAT or the MTs, i.e. $max[\Omega_{rad,p}, \Omega_{mech, p}]$. 

 It is important to keep in mind the important assumption in Table 1.   The change of the direction of grain alignment as a result of  inefficient internal relaxation happens assuming that the rotation at low-$J$ attractor points is stabilized by the external pinwheel torques (Purcell 1979). If such torques are inefficient, we expect the "wrong" alignment with the long grain axes parallel to AA to be unstable. The effects of gaseous bombardment on this type of alignment have not been explored in Hoang \& Lazarian (2009), but we can expect that grains are going to wobble significantly, spending a significant amount of time with the grain axes perpendicular to the AA. The situation is even more uncertain with respect to METs. The "wrong" alignment arising from the carbonaceous precessing quickly about an electric field therefore is a more robust effect.  

The new element in Table 1 is that $\Omega_k$ is compared not only with $\Omega_B$, but also with $\Omega_E$, which can be significantly larger than $\Omega_B$. Thus, if naive application of the LH07 results would entail the conclusion that for $\Omega_k>\Omega_B$ the RAT alignment happens with respect to ${\bf k}$,
if $\Omega_E>\Omega_k$ the grains will be aligned still with respect to ${\bf B}$, as ${\bf E}$ is perpendicular to ${\bf B}$. 

As we discussed earlier, the alignment with the long axis along the magnetic field was predicted in Hoang \& Lazarian (2009) for grains aligned by RATs on the time scales less than the time of internal relaxation. The results in Table 1 extend the range of conditions when such alignment is possible. They also augment the picture with the effects of electric field ${\bf E}$ as well as the effects induced by the existence of high-J attractor points. Note, that the probability of their presence for realistic irregular grains became more clear only recently after the numerical study in Herranen et al. (2019).

We do not want to hide potential uncertainties related to our estimates of the precession rates obtained in \S \ref{sec:precess}. For instance, one can easily observe that Eqs. (\ref{rad_p}) and (\ref{Omegael}) testify that both $\Omega_{rad, p}$ and $\Omega_{E}$ are inversely proportional to the grain angular velocity, while the Larmor precession given by Eq. (\ref{prec}) does not change with the increase of angular velocity.  Taking into account that the magnetic moments of the carbonaceous grains can be enhanced by an order or magnitude (see W6) or more (see Lazarian \& Draine 2000), due to the presence of paramagnetic species in the form of free radicals, one can see that the for the precession of fast rotating grains the relation of the precession rates may become different, that is,
 $\Omega_{rad, p}<\Omega_{E}<\Omega_B$ for the typical interstellar radiation field.  

All in all, Table 1 testifies that the interpretation of polarization data may not be trivial for carbonaceous grains. At the same time, the detection of the difference of the alignment of carbonaceous and silicate grains or the differences of the polarization direction and the magnetic field direction inferred using velocity gradients (see Lazarian \& Yuen 2018)  can shed light on the nature of carbonaceous dust and the nature of the environment around dust grains.   

\begin{table*}[!htbp]
\renewcommand\arraystretch{1.5}
\centering
\begin{threeparttable}
\caption[]{Modes of carbonaceous grain alignment}\label{tab:com} 
  \begin{tabular}{c|c|c|c|c}
     \toprule
      \multirow{2}{*}{Conditions}        &            \multicolumn{2}{c|}{$\Omega_k < \text{max} [\Omega_E, \Omega_B]$}   &  \multicolumn{2}{c}{$\Omega_k > \text{max} [\Omega_E, \Omega_B]$}     \\
                      \cline{2-5}
                                                        &  $\perp B$  & $\| B$   & $\perp k$   & $\| k $ \\
                      \hline
         $\tau_\text{int} < \tau_\text{damp}$, $\tau > \tau_\text{damp}$   &        \multirow{2}{*}{$\Omega_B > \Omega_E$} & \multirow{2}{*}{$\Omega_B < \Omega_E$}      &  \multirow{2}{*}{applicable}  & \multirow{2}{*}{not applicable} \\
                     \cline{1-1}
         $\tau_\text{int} > \tau_\text{damp}$, $\tau \gg \tau_\text{damp, high-J}$     & & & \\            
                     \hline
         $\tau_\text{int} > \tau_\text{damp, low-J}$   &        \multirow{2}{*}{$\Omega_B < \Omega_E$} & \multirow{2}{*}{$\Omega_B > \Omega_E$}      &  \multirow{2}{*}{not applicable}  & \multirow{2}{*}{applicable} \\
                     \cline{1-1}
         $ \Omega_k^{-1} <\tau < \tau_\text{int}$     & & & \\            
     \bottomrule
    \end{tabular}
 \end{threeparttable}
\end{table*}

\subsection{Randomization of grains due to the variations of grain dipole moment does not happen}
\label{sec:rand}

Our discussion of the RAT and MET alignment of carbonaceous grains is based on the assumption that these alignment mechanisms are more efficient than the randomization of grain angular momentum ${\bf J}$.  As we discuss in Appendix D, the fluctuations of grain electric moment do not induce an additional randomization of interstellar grains. Therefore, both RAT and MET alignment of carbonaceous grains are expected to be efficient.

\section{Observational consequences of the predicted alignment}
\label{observ}

Let us consider a few cases for which our predictions of the alignment direction can be observationally tested. 

\subsection{Circumstellar accretion disks}
\label{circum}

Polarization from circumstellar disks has been studied both theoretically and observationally (see Cho \& Lazarian 2007, Hughes et al. 2009, Kataoka et al. 2016, Li et al. 2016, Tazaki et al. 2017, Bertrang et al. 2017). Our present study suggests that due to higher gas densities in such systems, $\tau_{rand}$ can be smaller than $\tau_{int}$, i.e. the alignment of carbonaceous grains there can happen in the regime of marginal internal relaxation. The degree of alignment is decreased, however, due to the decreased absolute values of RATs that the grains experience  for angle between the magnetic field and radiation direction $\psi=90$ degrees (see Figure 8 in Lazarian \& Hoang 2019). 

If $\Omega_E>\Omega_B$, the angle $\psi$ changes from 0 to 90 degrees as the charged grains performs gyrorotation with $\omega_{gyro}$ (see Eq. (\ref{gyro})). For this to happen $\omega_{gyro}^{-1}$ should be less than $\tau_{rand}$. This will entail the situation that a significant portion of grains to have only low-J attractor point (see Figure 24 in LH07). In this situation the direction of alignment will depend on whether $\tau_{rand}$ is larger or smaller than $\tau_{int}$. If $\tau_{int}<\tau_{rand}$, the alignment will happen with long grain axes perpendicular to ${\bf E}$ and therefore parallel to ${\bf B}$. In the opposite regime, i.e. $\tau_{int}<\tau_{rand}$ the alignment will happen with long axes parallel to
${\bf E}$ and therefore perpendicular to ${\bf B}$.

If $\Omega_k$ is larger than $max[\Omega_B, \Omega_{E}]$ the corresponding $\psi$ is 0 and Figure 24 in LH07 shows that for the distribution of $q^{max}$ from Herranen et al. (2019) most grains will have low-J attractor points only. In this setting grains will act differently depending whether $\tau_{ran}>\tau_{int}$ or $\tau_{ran}<\tau_{int}$. In the first case the alignment with axes parallel to radiation, i.e. in the radial direction, is expected. In the second case, the grain long axes will be perpendicular to the radial direction. 

\subsection{Envelopes of evolved stars}

Carbonaceous grains are are definitely present in the carbon-rich circumstellar envelopes of stars. The recent polarization study of such an environment for an asymptotic branch star by HAWC+ instrument on the SOFIA flying observatory revealed the alignment with grain axes parallel to the radiation direction (Andersson et al. 2020). 

Consider first $\Omega_k< \Omega_E$. If $\tau_{int}<\tau_{rand}$, in view of the present study, this result can be interpreted as the E-RAT alignment, that is, the alignment in respect to ${\bf E}$  with charged carbonaceous grains rotating around the radial magnetic field.  The RAT alignment of such grains is, as we discussed earlier, will provide the long grain axes to be aligned along the radial magnetic field. If $\tau_{int}>\tau_{rand}$ the observed polarization can be explained as E-RAT alignment with respect to the toroidal magnetic field provided that most grains have only low-J attractor points. We find the latter option less likely than the former one.

In the case of radiation determining the axis of maximal precession, the alternative idea was discussed in Andersson et al. (2020) for the case of  the alignment of carbonaceous dust grains with the suppressed internal relaxation. Such grains can be aligned by the FA process discussed in \S \ref{FA} for any rate of internal alignment or by the IA process discussed in \S \ref{IA}, provided that $\tau_{int}>\tau_{rand}$ but most of grains have only low-J attractors. 

We note that the alternative process of {\it stochastic} mechanical alignment, for example, Gold (1952) alignment, is unlikely. Indeed, we expect the METs to dominate unless the grains are absolutely symmetric, which is highly improbable. The MET alignment acts similar to the RAT alignment we discussed above. Further studies should clarify which alignment process is acting in the particular environment.

\subsection{Cometary dust}

Cometary dust alignment presents another case when E-RAT alignment can be explored. In Lazarian (2007) and Hoang \& Lazarian (2014) the alignment with respect to the electric field of cometary coma was considered in order to explain the observed circular polarization within the single scattering mechanism.  It was assumed that interplanetary magnetic field cannot approach the surface of the comets. However, simulations by Rubin et al. (2014) revealed that this is not true: the interplanetary magnetic field can approach the surface of comets to a distance of 100 km or less. This motivated the suggestion in Kolokolova et al. (2016 ) that the alignment of comet dust with respect to the interplanetary magnetic field can be the cause of the observed circular polarization. 

In view of our current paper, we would like to point out to the evidence that comet dust contains pure carbonaceous particles (see Colangeli et al. 1989, Fomenkova et al. 1994). These particles are expected to move with high velocities, for example, more than 100 km/s, with respect to the interplanetary magnetic field, and this should result in the strong electric field that is perpendicular to ${\bf B}$. This opens a way to test the predicted effects of carbonaceous grain alignment observing circular polarization from comets.

\subsection{Time-dependent alignment in the vicinity of variable stars, novae and supernovae}

The time dependence of the alignment can be another important way to test our predictions. The FA (see \S \ref{FA}) is expected to take place for all $\sim 10^{-5}$~cm carbonaceous grains in the presence of sufficiently strong radiation sources. Similarly, the FA can happen as grains move in respect to the gas due to shocks.

The time-dependent alignment is easiest to observe in the vicinity of the radiation sources of changing intensity. Variable stars, novae, and supernovae are the examples of such sources (see Palet et al. 2015). 
Apart from the FA, one can expect to observe a slower change of the alignment arising from the IA (see \ref{IA}). For instance, as described in \S \ref{circum} the 90 degree change of the carbonaceous grain alignment can take place in the vicinity of the radiation source over the time $\sim \tau_{rand}$. These changes can be detected. We note that other processes that induce a time-dependent change of alignment near variable sources are discussed in Lazarian \& Hoang (2019). 

\subsection{Alignment of carbonaceous nanoparticles}
\label{nano}

Polycyclic aromatic hydrocarbons (PAHs) present a well-accepted example of carbonaceous nanoparticles existing in the ISM (see Draine 1989). 

\subsubsection{Resonance paramagnetic relaxation}

This paper deals with the bimodal nature of the alignment of carbonaceous grains $\sim 10^{-5}$~cm. 
PAHs can also be aligned. In Lazarian \& Draine (2000) the process termed resonance paramagnetic relaxation (RPR) has been introduced as a way of aligning carbonaceous nanoparticles (see also Hoang et al. 2014). The difference of the process from the traditional Davis-Greenstein (1951, henceforth DG, see also Roberge \& Lazarian 1999) paramagnetic relaxation is that the RPR takes place within the material that is magnetized so it has resonance paramagnetic response. The effective magnetization in RPR is provided by the Barnett effect that takes place within the material of the rapidly rotating grain. While the DG mechanism fails for rapidly rotating grains, the relaxation by the RPR is ensured by the effective Barnett magnetization, which is exactly the magnetization required to have paramagnetic resonance.\footnote{For small grains rotating with velocity $\omega$ there exist a problem of direct transfer of the energy $\hbar \omega$ associated with the flip of the spin to the grain lattice. The latter has much higher minimal vibrational frequency (see Cyvin 1982). This problem of the suppressed spin-lattice relaxation was addressed by Lazarian \& Draine (2000) by appealing to the indirect energy transfer via the Raman scattering of phonons (see Waller 1932). The corresponding estimates obtained by  Lazarian \& Draine (2000) were found consistent with the model of the RPR for nanoparticles. On the contrary, the calculations in Draine \& Hensley (2016) of the rate of the direct energy transfer from the spin system to grain lattice show that the direct process is inefficient for the nanoparticles.} It is obvious from Eq. (\ref{aleph_carb}) that for the grains smaller than $10^{-7}$ cm the $\Omega_B>\Omega_E$ and the alignment is to happen with respect to the magnetic field. In fact, in most ISM settings with also expect $\Omega_B> \Omega_{E}$ and therefore the predictions of the alignment in Lazarian \& Draine (2000) are not altered. 

\subsubsection{Cross section alignment}

Another alignment process is based on the interaction of carbonaceous nanoparticles with radiation. The RATs are very weak for the nano-particles, but another process of "cross-sectional" alignment can be efficient. This alignment was first introduced in Lazarian (1995) for a flow of atoms impinging a grain. The grains with different cross-sections exposed to the flow were shown to have different rate of residence at a given position. In Hoang \& Lazarian (2015) the mechanism was shown to be applicable to the carbonaceous nanoparticles that are exposed to the UV radiation flux. The absorption of individual photons by a nanoparticle induces its heating to a high temperature. The subsequent emission of photons induces diffusion the grain angular momentum. As the rate of photon absorption depends on the grain cross-section, the grains get aligned in the position for which their cross-section in respect to the photon flux is minimal. The calculations performed in Hoang \& Lazarian (2015) demonstrate that the polarization  degree arising of emission arising from nanoparticles, e.g. PAHs, in the reflection nebulae can be 5\% at 10~GHz and up to 20\% at 100~GHz.  The alignment is expected long grain axes parallel to magnetic field for the angle between the radiation and the magnetic field $\psi$ in the range from 0 to 45 degrees and perpendicular to magnetic field otherwise.
This emission can explain the polarization of 11.3 $\mu$m feature from MWC 1080 nebulae reported in Zhang et al. 2017. Due to fast grain rotation,  $\Omega_B>\Omega_E$ and the predictions in Hoang \& Lazarian (2015) are not altered by our present study.

\subsubsection{Mechanical alignment in shocks}

While the RATs get inefficient as the grain size becomes much smaller than the wavelength, there is no such a impediment for the mechanical alignment. Therefore MET alignment should be efficient for irregular nanoparticles and be more efficient that the traditional stochastic processes of mechanical alignment, e.g. Gold (1951) alignment. The question that requires further studies is whether at these scales the nanoparticles are irregular. The detection of alignment of such particles in shocks may provide an answer to this question. The expected MT alignment is with long grain axes perpendicular to the flow, while the Gold (1951) and similar stochastic mechanical processes (see Lazarian 1994, 1995ab) provide the alignment that is different 90 degrees to the MET alignment.

\section{Discussion}
\label{sec:dis}

The detection and study of the predicted alignment of carbonaceous dust is important for understanding of carbonaceous dust physics, evolution of dust as well as physical enviroment that the grains are experiencing. Potentially such studies can help to better accounting for the galactic foreground polarization that interferes with the studies of polarization of cosmological origin.

\subsection{Carbonaceous grains in astrophysical environments}

The widely accepted models of interstellar grains assume the existence of separate populations of silicate and carbonaceous particles (see Draine \& Li 1984). While condensation of materials in the form of grains happen in the stellar outflows the survival time of these stardust grains is the interstellar medium is about a few million years (see Draine 2011). Thus most of the interstellar dust is reprocessed and produced in the interstellar medium with observations constraining the accepted models of grains (see Draine 2003, Hensley \& Draine 2020). 

The aliphatic CH feature at 3.4$\mu$m originating from the surface of carbaceous grains exposed to UV light and atomic hydrogen (Chiar et al. 2013), does not show  polarization  (Chiar  et  al.  2006). From the point of view of RAT theory, the absence of the 3.4$\mu$m in diffuse media can be explained if carbonaceous grains are significantly smaller than the typical wavelength of the radiation that they experience. Both the wavelength of the radiation and the size of the grains should, however, vary in the astrophysical conditions. Therefore the RAT alignment of at least some fraction of carbonaceous grains is inevitable. In addition, MET alignment, e.g. in shocks, is less dependent on the grain size and it should also happen. 

\subsection{Alignment complexity revealed by this study}

This study reveals the complexity of carbonaceous alignment that can decrease the degree of the expected polarization and complicate the observed polarization pattern. The grain alignment theory developed for the silicate grains predicts that $10^{-5}$~cm grains can be aligned either with their axis of maximal moment of inertial either parallel to $\bf B$ or parallel to ${\bf k}$, depending on whether $\Omega_B$ or $\Omega_k$ are larger. In practice this means that the long grain axes are perpendicular to the alignment axis. 

The situation is more complicated for carbonaceous grains. For these grains one has to account for an additional precession rate, namely, the rate of grain precession in electric field $\Omega_E$. If $\Omega_E>\Omega_B$ the alignment happens with grain long axes parallel to magnetic field, making the interpretation of polarization measurements more complex. Indeed, $\Omega_E$ depends on the turbulent acceperation of grains, grain charging and grain electric dipole moment, all of these chaging with the grain composition and grain environment. 

Our estimates in \S \ref{sec:dyn} are suggestive that the alignment of $\sim 10^{-5}$~cm grains in realistically turbulent interstellar environments tends to be with long grain axes {\it parallel} to the magnetic field ${\bf B}$. This conclusion, however, can be altered if grains experience strong suprathermal torques (see \S \ref{supra}) that can change the relation between the precession rates making $\Omega_B>\Omega_{E}$. In this situation the grain alignmnet changes 90 degrees, i.e. grains get aligned  with long axes {\it perpendicular} to ${\bf B}$.

Depending on the radiation field, grains may have $\Omega_B<\Omega_E$ for the low-$J$ attractor point the opposite relation for a high-$J$ attractor point. Thus, depending on the angle between the magnetic field and the radiation direction $\psi$ some carbonaceous grains can be aligned with long grain axes parallel to {\bf B}, while grains at low-$J$ can be aligned in a perpendicular direction. 

In addition, even if grains are low-$J$ attractor points, the gaseous bombadment can occasionally move them to the vicinity of the high-$J$ stationary point, which is a repellor point. During this excursion, the grain angular momentum will increase and, if the increase is significant enough $\Omega_B$ can become larger than $\Omega_E$, inducing a complicated time-dependent alignment pattern.

The same change of alignment and therefore of the polarization direction happens if carbonaceous grains have significantly larger number density of free radicals (see Lazarian \& Draine 2000). In addition, if some of the $\sim 10^{-5}$ grains contain both silicate and carbonaceous fragments, their alignment is also {\it perpendicular} to ${\bf B}$ (see \S \ref{composite}).

Due to lower magnetic moment of carbonaceous grains the k-alignment in respect to the radiation or mechanical flow direction is more likely. This was argued in Lazarian \& Hoang (2019) as another reason why the expected polarization from carbonaceous grains can be weaker.\footnote{At the same time Lazarian \& Hoang (2019) made a mistake accepting the possibility that the anomalous randomization could be be a possible reason for a poor alignment of carbonaceous grains. As we showed in \S \ref{sec:rand} and Appendix D the variations of the electric moment of grains do not induce any additional randomization.}

The complexity of the alignment of carbonaceous grains may be an impediment for magnetic field tracing. However, the sensitivity of their alignment to the composition as well as environmental factors can be used successfully to probe the nature of dust and its environment with polarimetric means. Note, that in some cases the flip of the polarization in respect to magnetic field can be detected with using velocity gradients for tracing magnetic field direction (see Lazarian \& Yuen 2018).

We note, that the variations in the environment and properties of carbonaceous grains can induce the variations of the polarization direction along the line of sight and, as a result, significantly decrease the polarization observed at 3.4$\mu$m. This mitigates the puzzle of why we do not see the polarization in 3.4$\mu$m but does not solve it. We believe that if carbonaceous grains are present in the interstellar medium, molecular clouds and accretion disks, the signatures of the alignment should be observable and this paper should contribute to the corresponding searches. 

 Although it is not widely appreciated by the community, the alignment of large grains, e.g. grains with $a_{eff}> 10^{-4}$~cm can proceed with the axis of maximal moment of grain inertia being perpendicular to the alignment axis (Hoang \& Lazarian 2009b). For the RAT alignment it was shown that in the presence of the stabilizing pinwheel Purcell (1979) torques this alignment at low-J attractor points results in the grain alignment with long grain parallel to the alignment axis. This effect is the consequence of the internal alignment getting inefficient for both sufficiently large silicate and carbonaceous grains. Such big grains are expected to be present in circumstellar accretion disks, for instance. The resulting "wrong" alignment of big grains is by itself a complication for interpreting the polarization measurements of silicate grains. For the carbonaceous grains, the interpretation the interpretation is further complicated by the existence of the an additional parameter, namely, the precession rate $\Omega_E$ (see Table 1). This further stresses the importance of spectro-polarimetric measurements as a way of probing both grain properties and grain environment. 

\subsection{Segregation of silicate and carbonaceous fragments and the alignment of composite grains}
\label{composite}

The issue whether carbonaceous and silicate materials are segregated is unclear by itself. It was argued in Lazarian \& Hoang (2019) that the separation of grains of different composition can be achieved by the action of RAT centrifugal disruption introduced discussed in Hoang (2017). Lazarian \& Hoang (2019) speculated that this can happen if bonds between the fragments of carbonaceous and silicate grains are weaker than between the silicate \& silicate and carbonaceous \& carbonaceous fragments. 

While this process can be feasible in the diffuse interstellar medium it is not likely segregate silicate and carbonaceous fragments in molecular clouds where the radiation field is reduced. Therefore Lazarian \& Hoang (2019) concluded that the grains in molecular clouds contain both silicate and carbonaceous fragments. Such composite grains should have magnetic moment orders of magnitude larger than pure carbonaceous grains. Therefore the Larmor frequency of the composite grains  $\Omega_B$ is   expected to be larger than $\Omega_{E}$ (see Eqs. 
(\ref{prec}) and (\ref{Omegaelf}) and the alignment of composite grains is expected to be similar to the silicate ones.\footnote{In molecular clouds we also expect magnetic inclusion to be incorporated in the composite grains. The latter further increases the efficiency of RAT alignment (Lazarian \& Hoang 2008, Hoang \& Lazarian 2016).} 

According to Giles Novak (private communication) the the far infrared observations testify that carbonaceous grains are aligned in molecular clouds. In view of our discussion above, this can be the consequence of grains in molecular clouds having composite silicate-carbonaceous nature.

\subsection{Comparison with silicate grain alignment}

Our study shows that carbonaceous grains with sizes larger than $\sim 10^{-5}$ cm are expected to be aligned, but the polarization arising from such grains in typical interstellar conditions is perpendicular to the polarization arising from their silicate or mixed carbonaceous-silicate counterparts. The alignment of the carbonaceous grains can become similar to the silicate grains if either of the following conditions is applicable: (a) the magnetic fields locally is orders of magnitude larger than on average in the interstellar medium, which is unlikely in the ISM, but possible in some molecular clouds, (b) the level of turbulence is locally is significantly lower than on average in the ISM, which can be true over a relatively small fraction of the volume, (c) the rotational velocities of carbonaceous grains exceed significantly our estimates, (d) the concentration of free radicals and other species with uncompensated spins is orders of magnitude larger compared to our estimates in the paper. The latter point does not have much constraints at the moment due to our ignorance of the detailed properties of the interstellar grains. Therefore the studies of polarization arising from carbonaceous grains can provide important insights bot into the interstellar grain composition and their dynamics.

The physics of the alignment that we discussed is also applicable to silicate grains. However, larger magnetic moment of silicate grains makes their E-alignment only possible for much stronger electric fields, limiting the parameter space for which the effect is observable. Similarly, a faster internal relaxation makes the alignment less feasible. If silicate grains have enhanced magnetic response (see Lazarian \& Hoang 2019 and ref. therein) the effects discussed in this paper are further mitigated.

\subsection{Dealing with complexity}

The history of grain alignment theory traces the increasing complexity of the processes that one has to deal with. Instead of spherical grains that the Davis-Greenstein (1951) or prolate/oblate grains that the Gold (1952) mechanisms dealt with, the RAT and MT alignment mechanisms necessarily deal with realistic irregular grains. This step allowed to both identify much stronger alignment processes and get the understanding that both the alignment in respect to ${\bf B}$ and ${\bf k}$ are feasible (see LH07). This paper adds an additional element to the grain dynamics complexity, namely, the ambient turbulence. The turbulence both induces fast motion of grains in respect to ${\bf B}$ and its compression of magnetic field also create the electric field ${\bf E}$. This results in a new type $E$-type RAT or MT grain alignment the properties of which we are just starting to explore. Below we outline a few yet unclear issues related to the process. 

 For the grains accelerated by turbulence the relative gas-grain direction of motion depends on the particular mechanism of acceleration. For instance the electrric field acting during gyroresonance (see Yan \& Lazarian 2003) tend to increase the perpendicular to magnetic field component of velocity and therefore it tend to be perpendicular to magnetic field. At the same time, in the process of  Transient Time Damping (TTD) acceleration (Hoang et al. 2012, Xu \& Lazarian 2018) it is the parallel to magnetic field velocity increases.\footnote{The process of the TTD acceleration is based on grain surfing  the fronts of magnetic compressions. Incidentally, the accelerating electric field in this case is {\it parallel} to the mean magnetic field. This may further complicate the alignment of carbonaceous grains.} As a result, an additional component of electric field along the magnetic field is present in the TTD acceleration. We do not expect this to change significantly the direction of grain precession if the grains have significant velocities perpendicular to magnetic field. Nevertheless, the issue requires further studies.

  In shocks the typical grain velocities can be larger than the the averaged velocities in turbulent medium. If the magnetic field is compressed in the shock and the grain is crossing the shock, the electric field acting upon the grain will be in the plane of the shock and perpendicular to the magnetic field direction. However, if the time of the passing of the grain through the compressed magnetic field is shorter than the grain precession period given by Eq. (\ref{t_mex})  the alignment can happen in respect to the direction of the grain motion in respect to gas.
  
  An additional unexplored issue is the  dissipation of energy of a rotating grain in the external electric field ${\bf E}$. This effect may be important for some of the materials with electric domain structure, e.g. ferroelectrics (see Werner 1957) and it can induce dissipation aligning grains to rotate with {\bf J} parallel to ${\bf E}$.\footnote{Depending on the grain composition, more exotic effects, e.g. magnetoelectric effect, inducing the response of in terms electric polarization to magnetization and vice versa can be important (see Dzyaloshinsky 1960). The effect is observed, for instance, in Cr$_2$O$_3$, but we are not aware of a similar effect in carbonaceous materials.  We also expect this effect to be strongest in respect to Barnett magnetization, which is weak for carbonaceous grains. Therefore we do not discuss this effect in this paper.}

\section{Summary}
\label{sec:sum}

Polarization arising from aligned grains is routinely used to study magnetic fields in various astrophysical environments, e.g. interstellar medium, molecular clouds, accretion discs, circumstellar regions. However, polarization studies can also be useful in understanding the chemical composition of grains, provided that we better understand the processes of grain alignment. In this paper we studied the alignment of carbonaceous grains in order to get insight to a number of important astrophysical questions, e.g. (a) whether the difference of alignment of silicate and carbonaceous grains can account for the failure of measuring 3.4 $\mu$m polarized feature in the diffuse interstellar medium spectra, (b) whether we expect to see  carbonaceous material aligned in molecular clouds and circumstellar regions, (c) what are the expected properties of the polarization arising from carbonaceous grains. The results of our study above can be briefly summarized in the following way.

\begin{itemize}
\item In the presence of interstellar radiation field it is not feasible to avoid RAT alignment of carbonaceous grains with sizes $10^{-5}$ cm and larger. The earlier reported process of efficient randomization of carbonaceous grains due to thermal flipping and variations of grain charge is invalid.

\item We demonstrated that in realistically turbulent medium the carbonaceous grains are expected to be aligned with the direction of electric ${\bf E}\sim {\bf v}\times {\bf B}$ field that arises from the gyrorotation of the grains about magnetic field as well as electric field arising from the turbulent compressible motions of magnetized interstellar medium. This constitutes a new mode of grain alignment, ${\bf E}$-alignment. 

\item In the process of ${\bf E}$-alignment the carbonaceous grains align with long axes parallel to magnetic field, making the polarization arising from carbonaceous and silicate grains perpendicular to each other.

\item The alignment of composite silicate-carbonaceous grains, e.g. grains in molecular clouds, as well as carbonaceous grains with significantly enhanced concentration of free radicals and impurities with uncompensated electron spins is similar to the alignment of silicate grains.

\item Due to lower internal relaxation of energy, carbonaceous grains are more likely, compared to silicate grains, get aligned with their long axes parallel to the direction of the radiation for the RAT alignment and the direction of the flow for the MT alignment. 

\item The alignment of carbonaceous nanoparticles via resonance paramagnetic relaxation and cross-sectional mechanisms is taking place with long axes perpendicular to magnetic field, in agreement with earlier studies of the subject. 

\item The studies of grain alignment of dust of carbon-rich circumstellar regions provide a way to test the alignment that is predicted in this paper.
\end{itemize}

\acknowledgments
The NSF grant AST 1715754 and NASA TCAN AAG1967 grant are acknowledged.  We thank B-G Andersson for sharing the information on his unpublished results and Bruce Draine for illuminating discussions of the dust composition models. We also thank the anonymous referee for the helpful input. The Flatiron Institute is supported by the Simons Foundation.

\appendix 

\section{A: Dipole moment of carbonaceous grains}

The detailed calculations of grain dipole moment are performed in Jordan \& Weingartner (2009, henceforth JW09) and here we present some of the basic facts from that study. First of all, carbonaceous grains can have an intrinsic dipole moment as it is discussed in Draine \& Lazarain (1998). The nature of this dipole moment is related to the random attachment of polar molecules that constitute the carbonaceous grain. It was considered by Draine \& Lazarain (1998) that the attachment of new molecules results in the random walk process increase of the intrinsic dipole moment. Extrapolating this process to the $10^{-5}$~cm grains JW09 concluded that for typical ISM conditions the intrinsic dipole moment is not important compared to the dipole moment arising from grain charging. 

JW09 adopted the model of collisional and photoelectric charging from Weingartner \& Draine (2001) (see also \S \ref{charging}) as well as considered grain models with different conductivities. Grains composed of isolating materials were found to have higher dipole moments. The rate of dipole moment fluctuation in time plays an important role in W06 theory, but, as we discuss in the paper, it is not important for our study.

\section{B: Different regimes of RAT alignment}

\begin{figure}[htbp]
\centering   
   \includegraphics[width=9cm]{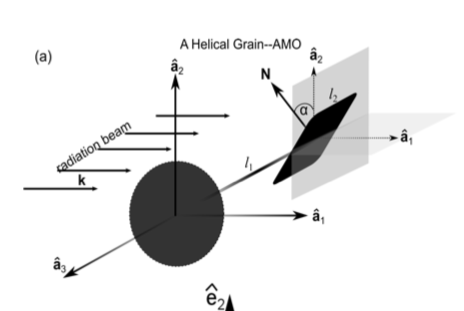}
\caption{LH07 model of a helical grain (AM0) consists of an oblate grain with a mirror attached to it at an angle $\alpha$. From Lazarian et al. (2015).}
\label{AMO}
\end{figure}

The present-day theory of RAT alignment is rather sophisticated. The basis of the understanding of RAT alignment physics is the Analytical Model (AMO) presented in LH07. This model consists of an ellipsoidal grain with a weightless mirror attached to it at 45 degrees as it is shown in Figure \ref{AMO}.

AMO allows the analytical description of its torques and, it was shown in LH07, in spite of its simplicity, it correctly describes the RATs acting on realistic irregular grains (see Herranen et al. (2019)). The helicity provided in AMO by the mirror corresponds to the actual helicity of irregular astrophysical grains as they interact with the radiation. It is the helicity that induces grain alignment.

LH07 showed that the alignment can happen in respect to magnetic field ${\bf B}$ or in respect to direction of incoming photon flow ${\bf k}$ and the axis of alignment is determined by the direction of the fastest precession, which can be either ${\bf B}$ and ${\bf k}$ (see \S \ref{sec:dyn}). Our present study adds to it the possibility of grain alignment in respect to electric field ${\bf E}$.

For the RATs to be efficient it is usually assumed that the condition that the wavelength of the radiation $\lambda$ is of the order of the grain size $a_{eff}$ (see Andersson et al. 2015). However, in many cases this is an oversimplification as it is evident from Eq. (\ref{eq:QRATs}). Indeed, a sufficiently strong flow of radiation with $\lambda> a_{eff}$ can produce an efficient alignment.

As we discuss in \S \ref{intern} the internal relaxation within $\sim 10^{-5}$ cm grains is fast. If grains rotate faster than the thermal rotation rate, this induces perfect alignment of grain angular momentum {\bf J} with the axis of the maximal moment of grain inertia {\bf a}. Thus the theory of RAT alignment was formulated in LH07 and subsequent publications (Hoang \& Lazarian 2008, 2009a) on the assumption that ${\bf J}$ and {\bf a} are perfectly aligned. For this setting, the alignment happens with long grain axes perpendicular to the AA. If this axis is the magnetic field, then grain get aligned with long axes perpendicular to the magnetic field. This coincides with the prediction of the paramagnetic alignment in Davis \& Greenstein (1951).

The RAT alignment happens at the attractor points which can correspond to RAT spinning up grains, i.e. high-$J$ attractor points, and the attractor points at which the RATs slow down grain rotation, i.e. the low-$J$ attractor points. Whether grain has a high-$J$ attractor point depends on the grain parameter $q^{max}$ introduced in LH07 as well as on the angle $\psi$ between the radiation anisotropy and the alignment axis (LH07). The grain alignment corresponding to the high-$J$ attractor points is perfect. 

At the same time, the low-$J$ attractor point alignment is unstable due to the low value of the angular momentum. First of all, as was shown in Lazarian (1994), the alignment of {\bf J} and {\bf a} significantly decreases in spite of internal relaxation being fast. This is a direct consequence of the Fluctuation-Dissipation theorem. 

Naturally, low, subthermal values of $J$ at the low-J attractor points make grain alignment unstable at these points. As a result, the gaseous bombardment easily randomizes grains at low-J attractor points. It was, however, shown in Hoang \& Lazarian (2009a) that in the presence of Purcell (1979) pinwheel torques the angular momentum of the grain can be increased at the low-$J$ attractor point and the alignment can be stabilized this way. The degree of alignment increases in this case.

As we discuss in \S \ref{intern} the internal alignment may be slower than the relevant RAT alignment processes. The triaxial grain in this situation can stably rotate both about the axis of maximal and minimal inertia. In this situation, it was shown in Hoang \& Lazarian (2009b) that the grains can be aligned at low-$J$ attractor point with long axes parallel to the AA. Similar to the earlier discussed case, this alignment is unstable and the grain is likely to leave the low-$J$ attractor, unless the rotation at that point is stabilized by Purcell (1979) torques. For the sake of simplicity, in this paper we assume that this is always the case. We should accept that the case of the RAT alignment in the absence of internal relaxation is a complex case which has not been studied in detail.

The RAT alignment can happen on time scales much less than the gaseous damping time. The time scale of the corresponding Fast Alignment (FA) is of the order of a precession period of a grain in the anisotropic radiation flux (LH07). It was shown in LH07 that during the FA it is most probable for a grain to move to the low-$J$ attractor point even for a given  $q^{max}$ and $\psi$ a high-$J$ attractor point exists.

\section{C: Mechanical torques on irregular grains}

The mechanical alignment of grains moving at high velocities in respect to the gas was pioneered by Gold (1952). That process and its modifications (see Lazarian 1995ab) are processes of stochastic alignment, however. A radically different process of alignment for irregular grains was introduced in 
Lazarian \& Hoang (2007b). 

The model of a grain in Figure \ref{AMO} was taken as the model of an irregular grain. The difference with the AMO is that the mechanical bombardment by atoms rather than the interaction with photons was employed. Lazarian \& Hoang (2007b) pointed out that for sufficiently favorable shapes of grains, the Mechanical Torques (METs) do not require supersonic relative velocities of grains and gas to align the grains. 

Numerical simulations of in Hoang et al. (2018) confirmed the higher efficiency of METs compared to the stochastic torques. This confirmed the MET alignment as the dominant process for among the mechanical grain alignment mechanisms. However, unlike the case of RATs, the AMO provides a qualitative, but not a quantitative representation of the torques acting on grains. This arises from the variations of the grain effective helicity as the grain interacts with the  corpuscular bombarding flux exposing its different facets.

We expect that the MET alignment to happen, in general, similar to the RAT alignment. For instance, the alignment of grains with efficient internal relaxation to happen with long grain axes perpendicular to the AA. In the absence of the internal relaxation the alignment can happen in low-$J$ attractor points with long grain axes parallel to the alignment axis. Similar to the case of RATs, this "wrong" alignment can be stabilized by pinwheel Purcell (1979) torques.

\section{D: Anomalous randomization}

 The anomalous randomization discussed in W06 potentially can challenge this conclusion. The process of anomalous randomization was a puzzling fact of grain alignment theory. For instance, in Lazarian \& Hoang (2019) we considered its effects and came to the conclusion that it presents a challenge for explaining any alignment of grains unless they have strongly enhanced magnetism. On the contrary, the authors of the grain alignment reviews in Lazarian et al. (2015) and Andersson et al. (2015) avoided the discussion of the process as they could not find a way to reconcile the effect with polarization observations. This motivates us to discuss this process in more detail. 

 The essence of the anomalous randomization is the change of the direction of ${\bf J}$ under the action of the time-variable torques that arise as the electric field acts on the fluctuating electric moment of the grain. The assumptions at the derivation of anomalous randomization are the following: a. magnetic field ${\bf B}$ provides the axis of precession; b. the torques induce a random walk for the angular momentum direction.

 However, in the paper above, we have shown that for typical ISM conditions  the precession of $10^{-5}$ carbonaceous grains the factor $\aleph$ is large and therefore the precession takes place in respect to the electric field ${\bf E}$. The magnetic field ${\bf B}$ only slightly alters the direction of a carbonacous grain precession. It is evident that the magnetic field does not directly interact with the electric dipole $p_{el}$ and thus its torque is constant. Therefore the presence magnetic field cannot randomize grains. 

The interaction of the electric field with the changing grain electric moment $p_{el}$ changes the torques acting on the grain, but the corresponding torques act in the plane perpendicular to the electric field. Therefore they can change the rate and direction of the precession of the angular momentum ${\bf J}$, but cannot change its angle with the alignment axis given by ${\bf E}$ (see \S \ref{sec:turb}).    

 It is important to notice that the torques arising from the changes of  $p_{el}$ do not introduce a random walk of ${\bf J}$. This is a consequence of the fact that $p_{el}$ fluctuates around its mean value. By itself, this precludes efficient randomization even if the parameter $\aleph$ given by Eq. (\ref{Omegaelf}) were small. A detailed quantitative discussion of the effects of the variations of electric torques arising from the variations of $p_{el}$ is provided in Lazarian \& Hoang (2020).\footnote{We note that the the variations of $p_{el}$ can be significant. For instance, large amplitude variations of $p_{el}$ arise from thermal flipping of grains arising from the coupling of the rotational and vibrational degrees of freedom. This effect introduced in Lazarian \& Draine (1999) and explored further in Weingartner (2009), Hoang \& Lazarian (2009) provides another way of flipping the direction of $p_{el}$. However, it is evident from what we discussed above, that this process does not randomize carbonaceous either.}



\begin{thebibliography}{55}
\expandafter\ifx\csname natexlab\endcsname\relax\def\natexlab#1{#1}\fi


\bibitem[Andersson et al.(2015)]{2015ARA&A..53..501A} Andersson, B.-G., Lazarian, A., \& Vaillancourt, J.~E.\ 2015, \araa, 53, 501

\bibitem[Andersson \& Potter(2007)]{2007ApJ...665..369A} Andersson, B.-G., \& Potter, S.~B.\ 2007, \apj, 665, 369

\bibitem[a(2015)]{03f} Andersson et al. 2020, preprint

\bibitem[Armstrong et al.(1995)]{Armstrong:1995} Armstrong, J.~W., Rickett, B.~J., \& Spangler, S.~R.\ 1995, \apj, 443, 209

\bibitem[Bertrang et al.(2017)]{2017MNRAS.464L..61B} Bertrang, G.~H.-M., Flock, M., \& Wolf, S.\ 2017, \mnras, 464, L61

\bibitem[Chepurnov \& Lazarian(2010)]{2010ApJ...710..853C} Chepurnov, A., \& Lazarian, A.\ 2010, \apj, 710, 853

\bibitem[Chiar et al.(2013)]{Chi2013}
Chiar, J. E., Tielens, A. G. G. M., Adamson, A. J., \& Ricca, A.2013, \apj, 770, 78

\bibitem[Chiar et al.(2006)]{Chi2006}
Chiar, J. E., Adamson, A. J., Whittet, D. C. B., et al. 2006,\apj, 651, 268

\bibitem[Cho \& Lazarian(2003)]{2003MNRAS.345..325C} Cho, J., \& Lazarian, A.\ 2003, \mnras, 345, 325

\bibitem[Cho \& Lazarian(2007)]{2007ApJ...669.1085C} Cho, J., \& Lazarian, A.\ 2007, \apj, 669, 1085

\bibitem[a(2015)]{03f} Collaboration, B.I.C.E.P., 2014. Detection of B-mode polarization at degree angular scales by BICEP2. Phys. Rev. Lett, 112, p.241101

\bibitem[a(2015)]{03f} Collaboration, P., Ade, P.A.R. and Aghanim, N., 2013. Planck 2015 results. XVI. Cosmological parameters, 1303, p.v1.
\bibitem[a(2015)]{03f} Colangeli, L., Schwehm, G.H., Bussoletti, E. 1989, \apj, 348, 718

\bibitem[Crutcher(2012)]{2012ARA&A..50...29C} Crutcher, R.~M.\ 2012, \araa, 50, 29

\bibitem[Cyvin(1982)]{1982JMoSt..79..423C} Cyvin, S.~J.\ 1982, Journal of Molecular Structure, 79, 423

\bibitem[Davis \& Greenstein(1951)]{1951ApJ...114..206D} Davis, L., \& Greenstein, J.~L.\ 1951, \apj, 114, 206

\bibitem[Das \& Weingartner(2016)]{2016MNRAS.457.1958D} Das, I., \& Weingartner, J.~C.\ 2016, \mnras, 457, 1958

\bibitem[Dolginov \& Mitrofanov(1976)]{1976Ap&SS..43..291D} Dolginov, A.~Z., \& Mitrofanov, I.~G.\ 1976, \apss, 43, 291

\bibitem[Draine(1989)]{1989IAUS..135..313D} Draine, B.\ 1989, Interstellar Dust, 313

\bibitem[Draine(2003)]{2003ARA&A..41..241D} Draine, B.~T.\ 2003, \araa, 41, 241

\bibitem[Draine(2011)]{2011piim.book.....D} Draine, B.~T.\ 2011, Physics of the Interstellar and Intergalactic Medium by Bruce T. Draine. Princeton University Press

\bibitem[Draine \& Hensley(2016)]{2016ApJ...831...59D} Draine, B.~T., \& Hensley, B.~S.\ 2016, \apj, 831, 59

\bibitem[Draine \& Lee(1984)]{1984ApJ...285...89D} Draine, B.~T., \& Lee, H.~M.\ 1984, \apj, 285, 89

\bibitem[Draine \& Lazarian(1998)]{1998ApJ...494L..19D} Draine, B.~T., \& Lazarian, A.\ 1998, \apjl, 494, L19

\bibitem[Draine \& Weingartner(1996)]{1996ApJ...470..551D} Draine, B.~T., \& Weingartner, J.~C.\ 1996, \apj, 470, 551

\bibitem[Draine \& Weingartner(1997)]{1997ApJ...480..633D} Draine, B.~T., \& Weingartner, J.~C.\ 1997, \apj, 480, 633

\bibitem[a(2015)]{03f} Dzyaloshinskii, I. 1960 Zh. Eksp. Teor. Fiz. 37: 881

\bibitem[a(2015)]{03f} Fomenkova, M.N., Chang, S.,  Mukhin, L.M. 1994, Geochmica em Cosmochimica Acta, 58, 4503

\bibitem[Gold(1952)]{1952MNRAS.112..215G} Gold, T.\ 1952, \mnras, 112, 215

\bibitem[Hensley \& Draine(2020)]{2020arXiv200202457H} Hensley, B.~S., \& Draine, B.~T.\ 2020, arXiv e-prints, arXiv:2002.02457

\bibitem[{Herranen {et~al.}(2019)Herranen, Lazarian, \&
  Hoang}]{Herranen:2019kj}
Herranen, J., Lazarian, A., \& Hoang, T. 2019, \apj, 878, 96

\bibitem[Herranen \& Lazarian(2020)]{2020arXiv200616563H} Herranen, J. \& Lazarian, A.\ 2020, arXiv:2006.16563

\bibitem[{Hoang(2017)}]{2017ApJ...836...13H}
Hoang, T. 2017, \apj, 836, 13

\bibitem[Hoang(2019)]{Hoang:2019} 
Hoang, T.\ 2019, \apj, 876, 13

\bibitem[{Hoang {et~al.}(2018{\natexlab{a}})Hoang, Cho, \&
  Lazarian}]{Hoangetal:2018}
Hoang, T., Cho, J., \& Lazarian, A. 2018{\natexlab{a}}, \apj, 852, 129

\bibitem[{Hoang {et~al.}(2010)Hoang, Draine, \& Lazarian}]{Hoang:2010jy}
Hoang, T., Draine, B.~T., \& Lazarian, A. 2010, \apj, 715, 1462

\bibitem[{Hoang \& Lazarian(2008)}]{HoangLazarian:2008}
Hoang, T., \& Lazarian, A. 2008, \mnras, 388, 117

\bibitem[{Hoang \& Lazarian(2009{\natexlab{a}})}]{2009ApJ...697.1316H}
Hoang, T., \& Lazarian, A. 2009{\natexlab{a}}, \apj, 697, 1316

\bibitem[{Hoang \& Lazarian(2009{\natexlab{b}})}]{2009ApJ...695.1457H}
Hoang, T., \& Lazarian, A. 2009{\natexlab{b}}, \apj, 695, 1457

\bibitem[{Hoang \& Lazarian(2014)}]{2014MNRAS.438..680H}
Hoang, T., \& Lazarian, A. 2014, \mnras, 438, 680

\bibitem[{Hoang \& Lazarian(2016{\natexlab{a}})}]{2016ApJ...831..159H}
Hoang, T., \& Lazarian, A. 2016{\natexlab{a}}, \apj, 831, 159

\bibitem[{Hoang \& Lazarian(2016{\natexlab{b}})}]{2016ApJ...821...91H}
Hoang, T., \& Lazarian, A. 2016{\natexlab{b}}, \apj, 821, 91

\bibitem[{Hoang \& Lazarian(2018)}]{Hoang:2018el}
Hoang, T., \& Lazarian, A. 2018, \apj, 860, 42

\bibitem[Hoang et al.(2012)]{2012ApJ...747...54H} Hoang, T., Lazarian, A., \& Schlickeiser, R.\ 2012, \apj, 747, 54

\bibitem[Hughes et al.(2009)]{2009ApJ...704.1204H} Hughes, A.~M., Wilner, D.~J., Cho, J., et al.\ 2009, \apj, 704, 1204

\bibitem[Jordan \& Weingartner(2009)]{2009MNRAS.400..536J} Jordan, M.~E. \& Weingartner, J.~C.\ 2009, \mnras, 400, 536 (JW09)

\bibitem[a(2015)]{03f} Esquinazi, P., \& Hohne, R. 2005, J. Magnetism Magnetic Materials, 290, 20

\bibitem[a(2015)]{03f} Kamionkowski, M. \& Kovetz, E.D., 2016.  ARA\&A, 54, pp.227-269.

\bibitem[Kolokolova et al.(2016)]{2016MNRAS.462S.422K} Kolokolova, L., Koenders, C., Goetz, C., et al.\ 2016, \mnras, 462, S422

\bibitem[Kowal et al.(2011)]{2011ApJ...735..102K} Kowal, G., de Gouveia Dal Pino, E.~M., \& Lazarian, A.\ 2011, \apj, 735, 102

\bibitem[Lazarian(1994)]{1994MNRAS.268..713L} Lazarian, A.\ 1994, \mnras, 268, 713


  \bibitem[{Lazarian(1995{\natexlab{a}})}]{1995MNRAS.277.1235L}
  
Lazarian, A. 1995{\natexlab{a}}, \mnras, 277, 1235
\bibitem[{Lazarian(1995{\natexlab{b}})}]{Lazarian:1995p3034}

Lazarian, A. 1995{\natexlab{b}}, \apj, 451, 660
\bibitem[{Lazarian(2007)}]{2007JQSRT.106..225L}
Lazarian, A. 2007, J. Quant. Spectrosc. Rad. Trans., 106, 225

\bibitem[{{Lazarian} {et~al.}(2015){Lazarian}, {Andersson}, \& {Hoang}}]{LAH15}
{Lazarian}, A., {Andersson}, B.-G., \& {Hoang}, T. 2015, in Polarimetry of
  stars and planetary systems, ed. L.~{Kolokolova}, J.~{Hough}, \& A.-C.
  {Levasseur-Regourd} ((New York: Cambridge Univ. Press)), 81

\bibitem[{Lazarian \& Draine(1999{\natexlab{a}})}]{1999ApJ...520L..67L}
Lazarian, A., \& Draine, B.~T. 1999{\natexlab{a}}, \apj, 520, L67

\bibitem[{Lazarian \& Draine(1999{\natexlab{b}})}]{1999ApJ...516L..37L}
Lazarian, A., \& Draine, B.~T. 1999{\natexlab{b}}, \apj, 516, L37

\bibitem[{Lazarian \& Draine(2000)}]{2000ApJ...536L..15L}
Lazarian, A., \& Draine, B.~T. 2000, \apj, 536, L15

\bibitem[{Lazarian \& Efroimsky(1999)}]{LazEfroim:1999}
Lazarian, A., \& Efroimsky, M. 1999, \mnras, 303, 673

\bibitem[Lazarian \& Hoang(2019)]{2019ApJ...883..122L} Lazarian, A., \& Hoang, T.\ 2019, \apj, 883, 122

\bibitem[{Lazarian \& Hoang(2007{\natexlab{a}})}]{2007MNRAS.378..910L}
Lazarian, A., \& Hoang, T. 2007{\natexlab{a}}, \mnras, 378, 910 (LH07)

\bibitem[{Lazarian \& Hoang(2007{\natexlab{b}})}]{LazarianHoang:2007b}
Lazarian, A., \& Hoang, T. 2007b {\natexlab{b}}, \apj, 669, L77

\bibitem[{Lazarian \& Hoang(2008)}]{Lazarian:2008fw}
Lazarian, A., \& Hoang, T. 2008, \apj, 676, L25


\bibitem[{Lazarian \& Yan(2002)}]{2002ApJ...566L.105L}
Lazarian, A., \& Yan, H. 2002, \apj, 566, L105

\bibitem[Lazarian \& Yuen(2018)]{2018ApJ...853...96L} Lazarian, A. \& Yuen, K.~H.\ 2018, \apj, 853, 96

\bibitem[Lazarian \& Vishniac(1999)]{1999ApJ...517..700L} Lazarian, A. \& Vishniac, E.~T.\ 1999, \apj, 517, 700

\bibitem[Li et al.(2016)]{2016ApJ...832...18L} Li, D., Pantin, E., Telesco, C.~M., et al.\ 2016, \apj, 832, 18

\bibitem[McKee \& Ostriker(2007)]{2007ARA&A..45..565M} McKee, C.~F., \& Ostriker, E.~C.\ 2007, \araa, 45, 565

\bibitem[Patat et al.(2015)]{2015A&A...577A..53P} Patat, F., Taubenberger, S., Cox, N.~L.~J., et al.\ 2015, \aap, 577, A53

\bibitem[P{\'e}rez et al.(2015)]{2015ApJ...813...41P} P{\'e}rez, L.~M., Chandler, C.~J., Isella, A., et al.\ 2015, \apj, 813, 41

\bibitem[{Purcell(1969)}]{Purcell:1969p3641}
Purcell, E.~M. 1969, Physica, 41, 100

\bibitem[{Purcell(1979)}]{Purcell:1979}
Purcell, E.~M. 1979, \apj, 231, 404

\bibitem[Roberge et al.(1995)]{1995ApJ...453..238R} Roberge, W.~G., Hanany, S., \& Messinger, D.~W.\ 1995, \apj, 453, 238

\bibitem[Roberge \& Lazarian(1999)]{1999MNRAS.305..615R} Roberge, W.~G., \& Lazarian, A.\ 1999, \mnras, 305, 615

\bibitem[Tazaki et al.(2017)]{2017ApJ...839...56T} Tazaki, R., Lazarian, A., \& Nomura, H.\ 2017,
\apj, 839, 56

\bibitem[Waller(1932)]{1932ZPhy...79..370W} Waller, I.\ 1932, Zeitschrift fur Physik, 79, 370

\bibitem[{Weingartner(2006)}]{2006ApJ...647..390W}
Weingartner, J.~C. 2006, \apj, 647, 390, (W06)
\bibliographystyle{apj.bst} 

\bibitem[Weingartner \& Draine(2001)]{2001ApJS..134..263W} Weingartner, J.~C. \& Draine, B.~T.\ 2001, \apjs, 134, 263

\bibitem[a(2015)]{03f} Werner, K., 1957 ``Ferroelectrics and Antiferroelectrics". In Frederick Seitz; T. P. Das; David Turnbull; E. L. Hahn (eds.). Solid State Physics. Academic Press. 

\bibitem[Xu \& Lazarian(2018)]{2018ApJ...868...36X} Xu, S., \& Lazarian, A.\ 2018, \apj, 868, 36

\bibitem[{Yan \& Lazarian(2003)}]{2003ApJ...592L..33Y}
Yan, H., \& Lazarian, A. 2003, \apj, 592, L33

\bibitem[{Yan {et~al.}(2004)Yan, Lazarian, \& Draine}]{Yan:2004ko}
Yan, H., Lazarian, A., \& Draine, B.~T. 2004, \apj, 616, 895

\bibitem[Zhang et al.(2017)]{2017ApJ...844....6Z} Zhang, H., Telesco, C.~M., Hoang, T., et al.\ 2017, \apj, 844, 6


\bibitem[Zhu et al.(2019)]{2019ApJ...882..135Z} Zhu, H., Slane, P., Raymond, J., et al.\ 2019, \apj, 882, 135



\end{thebibliography}
\end{document}